\documentclass[conference]{IEEEtran}
\usepackage[nocompress]{cite}
\usepackage{amsmath}
\usepackage{graphicx}
\usepackage{url}
\usepackage{booktabs}
\usepackage{graphicx}

\title{Seekable OCI: Lazy-Loading Container Images\\via Range-Request Indexing}

\author{\IEEEauthorblockN{James Thompson, Wayne Mesard, Jesse Butler, Sri Saran Balaji Vellore Rajakumar, Henry Wang}
\IEEEauthorblockA{Amazon Web Services}}

\begin{document}
\maketitle
\thispagestyle{plain}
\pagestyle{plain}

\begin{abstract}
Container image pulling accounts for the majority of pod startup time in Kubernetes environments.
Standard pull downloads the entire image before the container can start, even when the application accesses only a fraction of the image content at startup.
We present SOCI (Seekable OCI), a lazy-loading architecture that enables containers to start without downloading the full image.
SOCI builds an external index over standard OCI images, mapping files to byte ranges within compressed layers.
At runtime, a FUSE filesystem intercepts file accesses and serves them via HTTP range requests.
Unlike prior approaches that require image format conversion, SOCI works with unmodified images and standard registries.
The index is stored as an OCI referrer artifact, requiring no changes to images, registries, or deployment tooling.
On a 1.3\,GB Python web service image, SOCI reduces cold-start pull time from 20 seconds to approximately 2.8 seconds (7.4x speedup), with pull time independent of image size. Larger images see larger speedups (9.3x on a 2.5\,GB image) because SOCI pull time is constant while standard pull scales linearly.
We measure a crossover at 80\% access density: below this, lazy loading wins; above, parallel full pull is faster.
SOCI lazy loading is deployed in production on Amazon EKS and Amazon ECS Fargate (which launched 18.4 million tasks per day during Prime Day 2025~\cite{aws-prime-day-2025}), and has been serving lazy-load requests since 2023. EKS Auto Mode uses SOCI's parallel pull mode for GPU instances.
\end{abstract}

\section{Introduction}

Container image pulling accounts for most of container startup time~\cite{slacker-fast16}.
For a 1.3\,GB Python application image on an m5.xlarge instance pulling from ECR in the same region, a standard pull takes 20 seconds.
The container cannot serve its first request until the pull completes.
When a new pod launches, those 20 seconds are latency the end user experiences.

Launching a container image in the usual way is simple: containerd downloads every layer in the image, decompresses each one, and extracts the files onto disk.
Once all of this is done, the container application can start.
For a Python web service, the runtime needs approximately 80\,MB of interpreter, framework, and application code to begin serving requests.
The remaining 1.2\,GB (scientific libraries, optional dependencies, locale data, debug tools) sits untouched during initialization and might never be needed.
The container pays for 20 seconds of download but uses 5\% of what it downloaded.

It is natural and expected for container images to have more bytes than are strictly necessary.
Tools for slimming down images are useful, but require hard decisions about what is absolutely required and what is only needed occasionally.
Lazy loading container image data is a less sharp tool: mount the container filesystem immediately, and fetch file content on demand as the application accesses it.
This makes startup time proportional to the \emph{working set}, the bytes actually needed, rather than the total image size.
Our measurements confirm this: a lazy pull of the same 1.3\,GB image completes in approximately 2.8 seconds (7.4x faster). A 2.5\,GB image pulls in the same 2.8 seconds (9.3x faster), because SOCI pull time is constant while standard pull scales with image size.

Prior lazy-loading systems achieve this by converting images to new formats.
Slacker~\cite{slacker-fast16} demonstrated the value of lazy loading (5x startup reduction) using an NFS backend that required a custom storage layer.
eStargz~\cite{estargz} re-compresses each file in a tar layer as its own gzip member, so that file boundaries align with gzip member boundaries and individual files can be fetched independently.
Nydus~\cite{nydus} goes further, replacing the tar/gzip layer format entirely with a purpose-built filesystem image (RAFS) optimized for block-level deduplication and on-demand fetch.
Both eStargz and Nydus proved that lazy loading works in production.
But both require the image producer to convert images before pushing them, which changes the image digest, breaks existing signatures, and adds a conversion step to every build pipeline.

eStargz's design contains the seed of a simpler approach.
If the goal is to locate individual files within a compressed tar layer, you need a mapping from filenames to byte ranges.
eStargz embeds that mapping by restructuring the compression; SOCI provides it externally, as a separate artifact that references the unmodified image.

We present SOCI (Seekable OCI), a lazy-loading architecture that works with unmodified OCI images and standard container registries.
SOCI operates in two phases.
At build time, a one-pass scan of each compressed layer produces an index that maps files to byte ranges within the layer.
This index is stored as an OCI referrer artifact, associated with the image manifest but separate from the image itself; the image digest, layers, and signatures are unchanged.
At launch time, the container runtime discovers the index via the OCI referrers API, mounts a FUSE filesystem backed by the index, and fetches file content on demand via HTTP range requests as the application accesses it.

The key insight for SOCI and similar efforts is that we can extract individual uncompressed byte ranges and files from compressed archives if we have the right tools and artifacts.
Gzip-compressed tar archives consist of independent DEFLATE blocks; each block can be decompressed given its starting offset and a 32\,KiB decompression dictionary.
The zlib library has included helper functions for this (\texttt{inflatePrime()} and \texttt{inflateSetDictionary()}) since 2005, along with a reference implementation (\texttt{zran.c}~\cite{zranc}) demonstrating how to build a seek index over a gzip file.
This technique was not applied to containers until August 2022, when SOCI was open-sourced.
The Nydus project independently reached the same conclusion, adding a ``zran mode'' in November 2022~\cite{nydus-zran} that generates equivalent external indices over unmodified layers.
We trace some of this history in Appendix~\ref{sec:history}.

SOCI lazy loading is deployed in production on Amazon EKS (via optimized AMIs) and Amazon ECS Fargate (which launched 18.4 million tasks per day during Prime Day 2025~\cite{aws-prime-day-2025}), serving lazy-load requests since 2023.
EKS Auto Mode uses SOCI's parallel pull mode for GPU instances.
Our contributions are:

\begin{enumerate}
\item \textbf{External seekable index}: We show how to read individual files from a standard compressed OCI image without downloading or decompressing the full layer. The index is stored as an OCI referrer artifact alongside the image.

\item \textbf{FUSE-based lazy loading on cold pull}: We implement a containerd snapshotter that mounts container filesystems via FUSE on first encounter, fetching content on demand with sub-millisecond mount time. Layers already present in the content store are served from local cache without network access.

\item \textbf{Standard overlayfs on subsequent launches}: Once all layers are cached locally, SOCI assembles a standard overlayfs mount. The FUSE path is only active during cold pull; warm restarts have zero overhead versus a standard containerd deployment.

\item \textbf{Production deployment}: We characterize SOCI's behavior under degraded network conditions (packet loss, added latency, connection failure) and report on production deployment across Amazon EKS and ECS Fargate.

\item \textbf{Decision framework}: We characterize when lazy loading outperforms full download, establishing access density as the deciding variable. For dense-access workloads above the crossover, parallel full pull~\cite{soci-parallel-pull-blog} is a complementary approach.
\end{enumerate}

\section{Related Work}

\paragraph{Image format conversion.}
Slacker~\cite{slacker-fast16} demonstrated that lazy loading reduces container startup time by 5x using an NFS backend.
It requires a custom storage backend incompatible with standard registries.
eStargz~\cite{estargz} and stargz embed seek tables in a modified tar format, enabling random access without a separate index.
However, conversion changes the image digest, breaking content-addressable integrity and requiring rebuild or conversion pipelines.
Nydus~\cite{nydus} uses content-defined chunking and a custom filesystem driver (EROFS-based) for block-level deduplication and on-demand loading.
It achieves excellent performance but requires images to be converted to its native format.
DADI~\cite{dadi-atc20} proposes block-level image distribution with on-demand fetch, targeting VM images with extensions to containers.
All these systems require the image producer to opt in.
SOCI requires only the index to be built. Any party with read access to the image can build it after the fact.

\paragraph{Distribution optimization.}
Dragonfly~\cite{dragonfly} and Kraken~\cite{kraken} use peer-to-peer protocols to distribute image layers across cluster nodes, reducing registry load.
These are complementary to SOCI: peer-to-peer distribution can accelerate the range-request fetches that SOCI issues.

\paragraph{Image caching.}
Bottlerocket data volumes~\cite{bottlerocket} and Kubernetes DaemonSets can cache images on nodes before workloads are scheduled.
This eliminates pull time entirely for predictable workloads but requires advance knowledge of which images will be needed and consumes storage for cached images that may not be used.
SOCI addresses the unpredictable case: autoscale events, spot replacement, and new deployments where caching is not feasible.

\paragraph{Parallel full pull.}
The containerd community has proposed parallel layer downloads with chunked range requests~\cite{containerd-parallel}.
The parallel full-pull mode, integrated into the SOCI snapshotter, achieves near-linear bandwidth saturation for throughput-bound workloads~\cite{soci-parallel-pull-blog}.
This is complementary to lazy loading: when a workload needs the entire image, download it as fast as possible.
SOCI provides both modes (lazy loading for sparse access, parallel full pull for dense access), selectable per workload.

\paragraph{Serverless cold starts.}
SOCK~\cite{sock-atc18} and Catalyzer~\cite{catalyzer-asplos20} optimize serverless function cold starts through process caching and checkpoint-restore.
AWS Lambda SnapStart~\cite{snapstart} takes a similar approach, snapshotting an initialized runtime and restoring it on subsequent invocations.
These operate at the process level (reusing initialized runtimes) rather than the image level (avoiding download).
They are orthogonal to SOCI; both can be used together.

\section{Background and Motivation}

\subsection{OCI Image Format}

A container image conforming to the Open Container Initiative (OCI) Image Specification~\cite{oci-image-spec} consists of a manifest, a configuration object, and an ordered set of filesystem layers.
Each layer is a gzip-compressed tar archive representing a set of filesystem changes (additions, modifications, deletions) relative to the layer below it.
Layers are identified by the SHA-256 digest of their compressed content.

The image manifest lists layers in order with their compressed sizes.
The manifest provides no information about the \emph{internal structure} of layers: filenames, file sizes, and file offsets within the compressed archive are not exposed.
Without additional information, a client cannot determine which files are in a layer or where they are located without downloading and decompressing the entire layer.

\subsection{Container Image Pull Pipeline}

Imagine pulling a 1.3\,GB image with 10 layers from ECR onto an m5.xlarge instance in the same region.
Containerd executes the following:
\begin{enumerate}
\item \textbf{Manifest resolution}: Fetch the image manifest from the registry ($<$1\,s, $<$5\% of total).
\item \textbf{Layer download}: For each layer, issue an HTTP GET for the compressed blob, streaming through SHA-256 for integrity verification ($\sim$14\,s, 70\% of total).
\item \textbf{Unpack}: Decompress each layer (gzip) and extract files (tar) into the snapshotter's storage ($\sim$5\,s, 25\% of total).
\item \textbf{Snapshot preparation}: Stack layers via overlayfs to present the merged filesystem ($<$1\,s).
\end{enumerate}

Total: 20 seconds. The container cannot start until all four steps complete.

\subsection{How Images Get Bloated}

Container images accumulate content through layer stacking.
A typical Python web service image contains:
\begin{itemize}
\item OS base layer (debian/ubuntu): package manager, shells, utilities, locale data, man pages (50--100\,MB)
\item Language runtime: interpreter, standard library, pip (100--200\,MB)
\item Application dependencies: installed packages in site-packages (200\,MB--2\,GB)
\item Application code: the actual service ($<$10\,MB)
\end{itemize}

At startup, the application needs the interpreter, its direct dependencies, and its own code.
The OS utilities, locale data, dev headers, and unused transitive dependencies are accessed later or never.
They exist because images are built for generality (the same base image serves development, testing, and production) and because removing unused content requires active maintenance that most teams do not prioritize.

\subsection{Simple Model for Lazy Loading}

If a container accesses only $k$ bytes of an $n$-byte image at startup, downloading all $n$ bytes wastes $(n - k) / n$ of the transfer time.
For the measured Python Flask image: $k \approx 80$\,MB, $n = 1.3$\,GB, so 94\% of the download is wasted.

Lazy loading makes pull time proportional to $k$ rather than $n$.
The challenge is providing random access to files within compressed tar archives without modifying the archive itself.

\section{System Design}

\subsection{Design Principles}

SOCI was designed with the following constraints:
\begin{enumerate}
\item \textbf{No image modification.} The base OCI image must remain untouched. Preserving layer digests implicitly preserves anything depending on digests, like cryptographic signatures.
\item \textbf{No registry changes.} SOCI works with any registry that supports HTTP range requests (RFC 7233), which most registries do. This avoids the custom backend requirements of Slacker~\cite{slacker-fast16}.
\item \textbf{The container runtime decides.} The runtime chooses whether to lazy-load or fully download each layer, based on index availability, layer size, and potentially other signals. Kubernetes manifests, Helm charts, and CI/CD pipelines require no changes; adopting SOCI is a node-level configuration change.
\end{enumerate}

\subsection{The Seekable Index}

An OCI image layer is most commonly stored as a gzip-compressed tar archive.\footnote{The OCI Image Specification allows other compression formats (zstd, uncompressed), but gzip is dominant in practice. SOCI targets gzip layers; uncompressed layers need no lazy loading, and zstd support is future work.}
DEFLATE, the compression algorithm used by gzip, can decompress any block given that block's starting offset and the 32\,KiB sliding-window dictionary from the preceding block boundary.
Tar archives store files sequentially with fixed-size headers that record filename, size, and offset within the archive.

SOCI uses the tar headers and the 32\,KiB gzip decompression windows to build a seekable index, called a ztoc (``compressed table of contents'').
A ztoc has two sections: a block map recording the compressed offset and 32\,KiB decompression dictionary at each DEFLATE block boundary, and a file map recording each file's name, uncompressed offset, size, and which blocks contain its data.

Given the ztoc, serving an arbitrary file $F$ requires:
\begin{enumerate}
\item Look up $F$ in the ztoc to find its containing DEFLATE block(s).
\item Issue an HTTP range request for the entire compressed block(s) from the registry.
\item Initialize the decompressor with the stored 32\,KiB dictionary state for that block.
\item Decompress until reaching $F$'s offset within the uncompressed stream.
\item Return the decompressed bytes.
\end{enumerate}

The fetch granularity is the DEFLATE block (typically $\sim$128\,KiB compressed), not the individual file.
Any other files within the same block are available immediately after fetch, with no additional network request.

The index is generated by scanning the compressed layer.
Scanning a 470\,MB compressed layer takes 1--9 seconds depending on layer count (Section~\ref{sec:eval}).
The index size is proportional to the number of DEFLATE blocks (typically one entry per $\sim$128\,KiB of compressed data), yielding indices that are $\sim$1--3\% of the compressed image size.

\subsection{Index Storage}

SOCI has used two storage mechanisms for the index.
In v1, the index was stored as an OCI referrer artifact, discovered via the \texttt{referrers} API~\cite{oci-referrers}.
Since v2 (the default since v0.10.0), the index is bundled as a sibling descriptor in an OCI Image Index, with an annotation on the image manifest recording the index digest.
The v2 approach makes the image-to-index association immutable: it is part of the image artifact itself, rather than a mutable referrer that any party with write access could add or remove.

This design has three properties we rely on.
First, the image digest is unchanged because the index is a separate artifact, which means existing signatures remain valid.
Second, registries serve the index via standard OCI pull APIs.
Third, the index can be built and pushed in-line with the image build process, or independently after the fact.

\subsection{Runtime Architecture}

At runtime, SOCI operates as a containerd \emph{remote snapshotter} plugin.
The containerd snapshot API defines how container filesystems are prepared.
When containerd prepares a snapshot for a new container, it delegates to the configured snapshotter.

The SOCI snapshotter performs the following workflow for each container launch:
\begin{enumerate}
\item Receive a prepare request with the image reference from containerd.
\item Query the registry's referrers API for a SOCI index associated with the image manifest.
\item If an index exists and layers are not yet local, mount a FUSE filesystem presenting the merged layer view. File content is fetched on first access.
\item If layers are already local (warm restart), assemble a standard overlayfs mount from the content store.
\item If no index exists, fall back to standard behavior (download and unpack all layers, assemble overlayfs).
\end{enumerate}

Once a filesystem is available via any path, the container starts.

The FUSE filesystem presents the container's root filesystem by stacking layers in the appropriate order.
Directory listings and metadata queries are served directly from the ztoc (no network requests).
File reads trigger HTTP range requests to the registry for the compressed bytes containing the requested file, decompress them, and return the result.

\subsection{Two Fetch Paths}

SOCI pulls data on demand when the application needs it, and in the background to move all of the image bytes to the node's local store.
Both paths write fetched spans to a SOCI-specific local cache (separate from containerd's content store).
Once a full layer has been fetched, it is verified and committed to containerd's content store.

\paragraph{On-demand fetch.}
When the application reads a file that is not yet in the local span cache, SOCI fetches the minimum span(s) containing that file (one HTTP range request per $\sim$4\,MiB span).
The application blocks until the bytes arrive (measured median: 4.6\,ms per fetch on a warm connection to in-region ECR).
Subsequent reads of the same span are served from the local cache.

\paragraph{Background fetch.}
Concurrently, a background fetcher walks the remaining spans sequentially by offset and fetches them one at a time at lower priority than on-demand reads.
This is not speculative: it is required for availability.
We do not want any bytes in the container image to remain in the registry permanently; the goal is to make the full layer local within seconds to minutes of container start.
On-demand reads preempt background fetches, ensuring the application's immediate needs are always served first.
Once all spans are local, the layer is verified, committed to containerd's content store, and subsequent container starts use standard overlayfs with no FUSE involvement.

\paragraph{Parallel full pull.}
For workloads that access most of the image, a separate parallel full-pull mode~\cite{soci-parallel-pull-blog} saturates network bandwidth by downloading layers via concurrent range requests.

\subsection{Fallback Behavior}

SOCI degrades gracefully:
\begin{itemize}
\item If the registry does not support range requests (missing \texttt{Accept-Ranges} header), download the full layer and serve from local storage.
\item If the SOCI index is missing or corrupt, fall back to standard containerd unpack.
\item If a range request fails, retry with exponential backoff. After exhausting retries, download the full layer.
\end{itemize}

After the container starts, the filesystem appears available, with the only observable difference being latency.
Files not yet in the local cache require fetching from the registry on first access (measured median: 4.6\,ms per file on a warm connection).
Locally cached files have normal access times.

\section{Evaluation}
\label{sec:eval}

We are curious about four things.
First, how long does it take to generate a ztoc? Even though index generation is amortized over many container launches, it matters that this step is low-latency.
Second, how much does SOCI reduce pull time and launch time compared with the standard containerd overlayfs snapshotter?
Third, what is the effect of data access patterns? If the application reads most of the image, the per-file fetch overhead may exceed the time saved by deferring download, and a full pull becomes the better choice.
Fourth, do these results hold on managed compute (Fargate), where operators do not configure nodes?

\subsection{Experimental Setup}

\textbf{Infrastructure.} Amazon EC2 m5.xlarge instances (4 vCPUs, 16\,GB RAM, up to 10\,Gbps network) running Amazon Linux 2023 with containerd 1.7.27 and SOCI v0.14.0. All images stored in Amazon ECR within the same region (us-west-2).

\textbf{Methodology.} Each measurement uses a freshly provisioned containerd state (full data directory removal and service restart). Kernel page caches are dropped before each trial. Pulls use \texttt{nerdctl} with \texttt{--snapshotter soci}, which exercises the containerd transfer service and SOCI's lazy-mount path.

\textbf{Baseline.} Standard containerd pull with overlayfs snapshotter (single-connection download, sequential unpack).

\begin{table}[h]
\centering
\caption{Benchmark images used in evaluation}
\label{tab:images}
\footnotesize
\begin{tabular}{lrr}
\hline
Image & Compressed (MB) & Layers \\
\hline
ubuntu:22.04 & 28 & 1 \\
redis:7 & 42 & 7 \\
python:3.11-slim & 43 & 4 \\
debian:bookworm & 46 & 1 \\
nginx:1.25 & 68 & 7 \\
postgres:16 & 153 & 14 \\
golang:1.22 & 285 & 7 \\
node:18 & 377 & 8 \\
python:3.11 & 390 & 7 \\
python-flask & 493 & 10 \\
flask-bloated & 635 & 10 \\
\hline
\end{tabular}
\end{table}

\subsection{Index Generation Overhead}

\begin{table}[h]
\centering
\caption{SOCI index generation time across common container images}
\label{tab:e5}
\footnotesize
\begin{tabular}{lrrrr}
\hline
Image & Size & Max layer & Indexed & Gen \\
      & (MB) & (MB)      & layers  & (ms) \\
\hline
ubuntu:22.04 & 28 & 28 & 1 & 1,265 \\
redis:7 & 42 & 27 & 2 & 1,251 \\
python:3.11-slim & 43 & 28 & 2 & 1,357 \\
debian:bookworm & 46 & 46 & 1 & 1,858 \\
nginx:1.25 & 68 & 40 & 2 & 1,719 \\
postgres:16 & 153 & 108 & 2 & 4,417 \\
golang:1.22 & 285 & 88 & 5 & 3,897 \\
node:18 & 377 & 202 & 5 & 8,233 \\
python:3.11 & 390 & 225 & 5 & 8,992 \\
\hline
\end{tabular}
\end{table}

Index generation is a one-time cost per image version.
Table~\ref{tab:e5} shows generation time correlates with the number of indexed layers and their sizes, not total image size.
Postgres (153\,MB, 14 layers) generates in 4.4\,s because only 2 layers exceed the 10\,MB indexing threshold; python:3.11 (390\,MB, 7 layers) takes 9\,s because 5 layers are indexed.
In production, index generation runs asynchronously after image push and completes before the image is scheduled to any node.

Layers below 10\,MB are skipped (not indexed) because the overhead of range-request fetching exceeds the cost of downloading them in full.
These small layers are fetched eagerly during the SOCI pull phase, contributing to the 2.8\,s pull time measured in Table~\ref{tab:e1}.

\subsection{Pull Time: SOCI vs Standard}

\begin{table}[h]
\centering
\caption{Pull time comparison (true cold start, SOCI v0.14, n=5 SOCI / n=3 full)}
\label{tab:e1}
\begin{tabular}{lrrr}
\hline
Image (size) & Standard (ms) & SOCI (ms) & Speedup \\
\hline
python-flask (1.3\,GB) & 19,551 $\pm$ 200 & 2,630 $\pm$ 193 & 7.4x \\
flask-bloated (2.5\,GB) & 25,426 $\pm$ 161 & 2,731 $\pm$ 130 & 9.3x \\
\hline
\end{tabular}
\end{table}

\begin{figure}[h]
\centering
\includegraphics[width=\columnwidth]{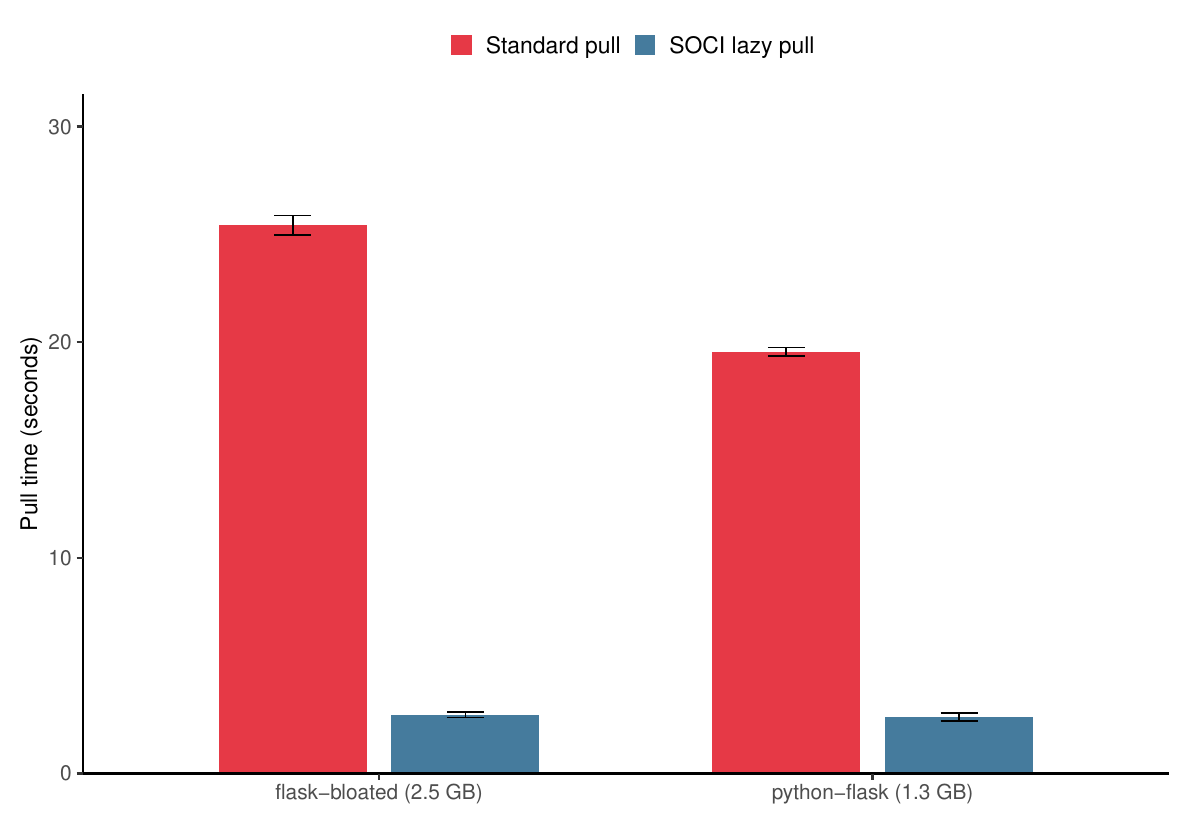}
\caption{Pull time: SOCI lazy pull vs standard containerd pull.}
\label{fig:pull}
\end{figure}

Table~\ref{tab:e1} shows pull time for two Python web service images of different sizes.
Each measurement uses a fresh cold start: both the SOCI daemon state and containerd content stores are wiped, kernel page caches dropped, and both daemons restarted.

The standard pull downloads all compressed layers and unpacks via tar extraction.
Pull time scales linearly with image size at approximately 10\,s/GB.

SOCI fetches only the image manifest, configuration, and SOCI index (approximately 10\% of the compressed image, or 42\,MB for a 409\,MB image).
Layer content remains in the registry; the FUSE filesystem fetches it on demand when the container accesses files.
SOCI pull time is constant at approximately 2.8\,s regardless of image size. The \textasciitilde300\,ms variation across runs reflects differences in SOCI index size (which scales with layer count) and network RTT jitter.

The speedup grows with image size because the SOCI pull time is constant while standard pull scales linearly: 7.4x for the 1.3\,GB image, 9.3x for the 2.5\,GB image.

Note: Table~\ref{tab:e1} measures \emph{pull time} only, the time from issuing the pull command to the image being mountable.
The 7.4x speedup is a pull-time speedup (9.3x for the larger image).
End-to-end speedup (pull plus startup to first useful work) depends on access density and is reported separately in Section~\ref{sec:crossover}.

\subsection{What SOCI Fetches at Pull Time}

To understand why SOCI pull time is constant, we instrumented the snapshotter daemon with bpftrace (tracing read syscalls) during 5 cold pulls of the python-flask image (493\,MB compressed).

\begin{table}[h]
\centering
\caption{Data transferred during pull: SOCI lazy vs standard (python-flask, 493\,MB compressed, n=5/3)}
\label{tab:e4}
\begin{tabular}{lrr}
\hline
Metric & SOCI & Standard \\
\hline
Bytes fetched & 51.2\,MB (10.9\%) & 493\,MB (100\%) \\
Network read operations & 7,799 & -- \\
Total pull time & 2,948 $\pm$ 69\,ms & 20,221 $\pm$ 318\,ms \\
\hline
\end{tabular}
\end{table}

At pull time, SOCI fetches approximately 11\% of the compressed image: the SOCI index artifacts (ztoc files containing DEFLATE block boundary maps), the image manifest and config, plus 4 small layers below the 10\,MB indexing threshold that are fetched eagerly rather than lazily.
No large-layer content is transferred.
The 2.9\,s pull time is dominated by HTTP round-trips for manifest resolution, index fetch, and small-layer download. Bulk data volume is not the bottleneck.
The remaining 89\% of image content stays in the registry until the container accesses it via FUSE at runtime.

\subsection{Access Density Crossover}
\label{sec:crossover}

SOCI lazy loading defers content fetching to runtime: files are fetched via HTTP range requests as the application accesses them.
This means SOCI's \emph{total} time (pull + startup) depends on how much of the image the application touches.
We measure this tradeoff using a synthetic 1\,GB image containing 1000 files, with a configurable entrypoint that reads a specified fraction (access density) before signaling readiness.

\begin{table}[h]
\centering
\caption{Total time (pull + startup) vs access density, synthetic 1\,GB image, n=10}
\label{tab:crossover}
\begin{tabular}{rrrr}
\hline
Access density & SOCI (ms) & Standard (ms) & Speedup \\
\hline
5\% & 3,401 $\pm$ 157 & 16,601 $\pm$ 71 & 4.9x \\
10\% & 4,401 $\pm$ 295 & 16,599 $\pm$ 49 & 3.8x \\
20\% & 5,932 $\pm$ 188 & 16,598 $\pm$ 45 & 2.8x \\
40\% & 9,202 $\pm$ 444 & 16,598 $\pm$ 1,933 & 1.8x \\
60\% & 12,990 $\pm$ 423 & 16,599 $\pm$ 50 & 1.3x \\
80\% & 17,043 $\pm$ 1,403 & 16,613 $\pm$ 85 & 1.0x \\
100\% & 20,089 $\pm$ 475 & 16,605 $\pm$ 854 & 0.8x \\
\hline
\end{tabular}
\end{table}

\begin{figure}[h]
\centering
\includegraphics[width=\columnwidth]{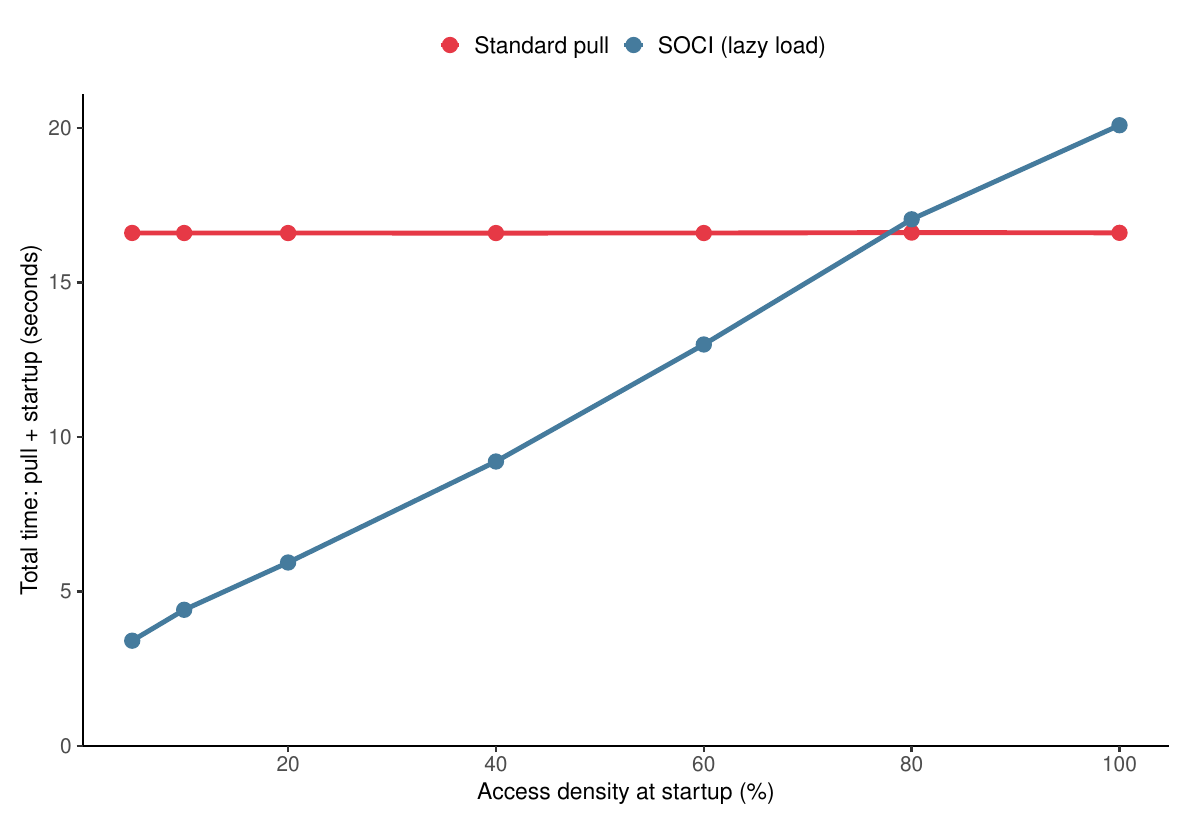}
\caption{Total time (pull + startup) vs access density. Crossover near 80\%: above this, standard pull is faster.}
\label{fig:crossover}
\end{figure}

Table~\ref{tab:crossover} compares lazy loading against \emph{standard single-stream pull}, which is containerd's default.
At 80\% density the two break even; below 80\%, lazy loading wins.
For workloads that access most of the image, parallel full pull~\cite{soci-parallel-pull-blog} is a complementary approach that saturates bandwidth to download the full image faster.

\subsection{Managed Compute: ECS Fargate}

To validate that SOCI's benefit extends to managed compute (where operators do not configure nodes), we measured pull time on Amazon ECS Fargate.
Fargate automatically detects SOCI indices and lazy-loads without configuration.

\begin{table}[h]
\centering
\caption{Fargate pull time with and without SOCI index, n=5}
\label{tab:fargate}
\begin{tabular}{lrrr}
\hline
Image & SOCI (ms) & Baseline (ms) & Speedup \\
\hline
python-flask (1.3\,GB) & 4,833 $\pm$ 1,075 & 19,740 $\pm$ 527 & 4.1x \\
flask-bloated (2.5\,GB) & 5,088 $\pm$ 1,297 & 25,150 $\pm$ 200 & 4.9x \\
\hline
\end{tabular}
\end{table}

\begin{figure}[h]
\centering
\includegraphics[width=\columnwidth]{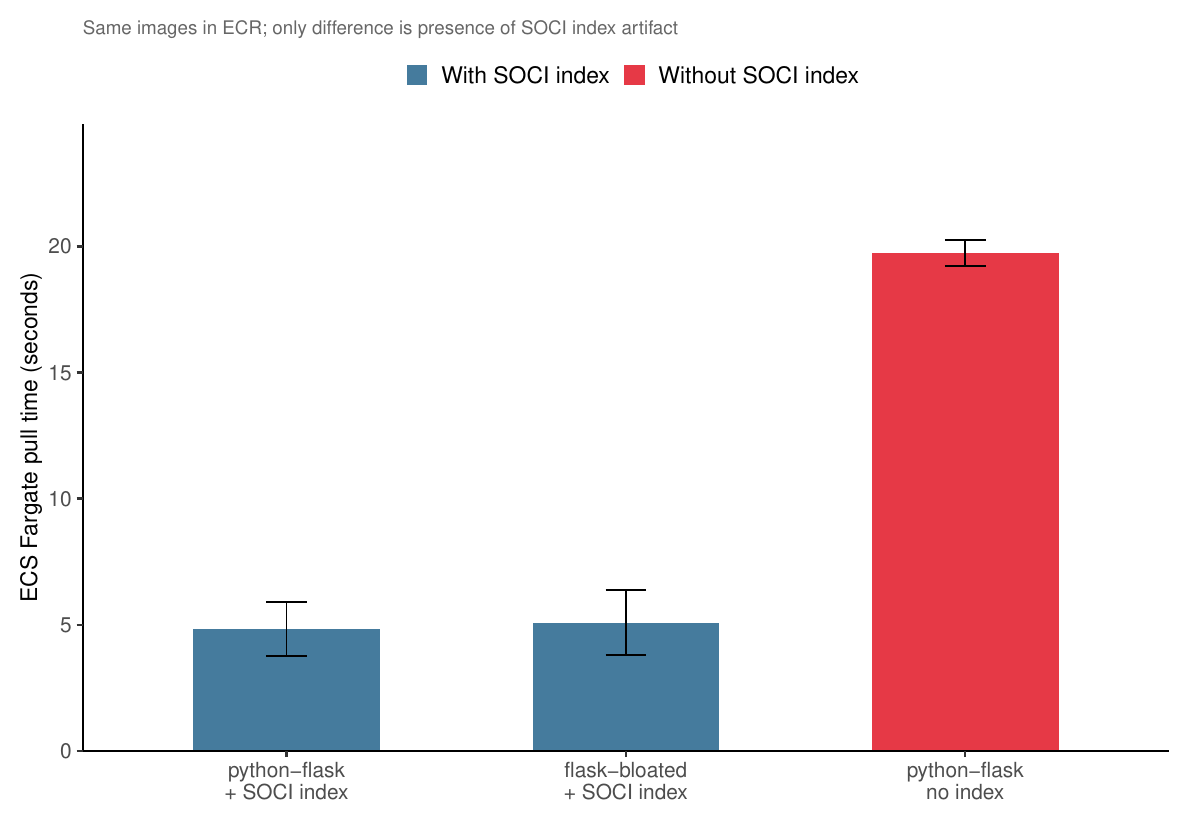}
\caption{ECS Fargate pull time for python-flask (1.3\,GB) and flask-bloated (2.5\,GB). Only difference is presence of SOCI index.}
\label{fig:fargate}
\end{figure}

Fargate achieves a 4x speedup with SOCI indices.
Pull time is again approximately constant regardless of image size (4.8\,s vs 5.1\,s).
The absolute pull time is higher than on EC2 (4.8\,s vs 2.8\,s) because Fargate's pull path includes additional task-provisioning overhead not present in bare containerd.

\subsection{Cost}

Consider 1000 cold starts of the python-flask image (1.3\,GB) on m5.xlarge (\$0.192/hr) in us-west-2.
This is a single data point, not a general formula; the numbers scale with image size and instance type.

The savings are idle compute time.
With standard pull, the instance runs for 20\,s per pull doing nothing useful (downloading and unpacking). With SOCI, that idle window shrinks to 2.8\,s.
Over 1000 pulls: \$1.07 of idle compute with standard pull, \$0.15 with SOCI.
The customer pays less because the instance starts doing useful work 17\,s earlier.

SOCI issues more HTTP requests to ECR (one range request per span vs one GET per layer).
ECR does not charge per-request for standard repositories; the only costs are storage (\$0.10/GB-mo) and data transfer (free in-region).
Even at raw S3 GET pricing (\$0.0004 per 1000 requests), the $\sim$123 additional range requests per pull would cost \$0.05 per 1000 pulls, roughly 5\% of the compute savings.
The SOCI index itself adds $\sim$1--3\% storage overhead.

\section{Discussion}

\subsection{When Lazy Loading Wins}

SOCI provides benefit on cold starts: whenever an image is not yet stored locally on a node, either because the node is new or the image is new.
Once an image is present on a node, both SOCI and standard containerd do the same thing (assemble an overlayfs mount from the content store) and complete in under 500\,ms (Section~\ref{sec:warm}).

For cold starts, our measurements on a 1\,GB synthetic image show SOCI provides a net end-to-end speedup at all access densities up to 80\%.
Lazy fetch overlaps with container execution: the container starts immediately and fetches files on demand, pipelining I/O with computation.
The speedup is largest at low access densities (4.8x end-to-end at 5\%) and decreases as more content is fetched at runtime.
The crossover occurs near 80\% access density; above this, the cumulative range-request overhead exceeds the time saved by deferring download.

SOCI has the biggest benefits on short-lived workloads, like serverless functions, batch jobs, or short-lived burst capacity. For these, the time to pull the entire image is a significant fraction of the total container lifetime.
A container that runs for 2 minutes spends 17\% of its life on a 20\,s pull; reducing that to 3\,s drops the overhead to 2.5\%.
A container that runs for 2 hours sees pull time shrink from 0.3\% to 0.04\%, real but negligible.
The shorter the workload, the more pull optimization matters.

\subsection{Warm Restart Behavior}
\label{sec:warm}

On a warm start, SOCI has the same performance as overlayfs.
Once an image is local to the containerd content store, subsequent pulls complete in $\sim$490\,ms, which is the time it takes to fetch a manifest from ECR in-region and verify it against the content store.
After verification, container startup takes $\sim$350\,ms for both SOCI and overlayfs.

\begin{figure}[h]
\centering
\includegraphics[width=\columnwidth]{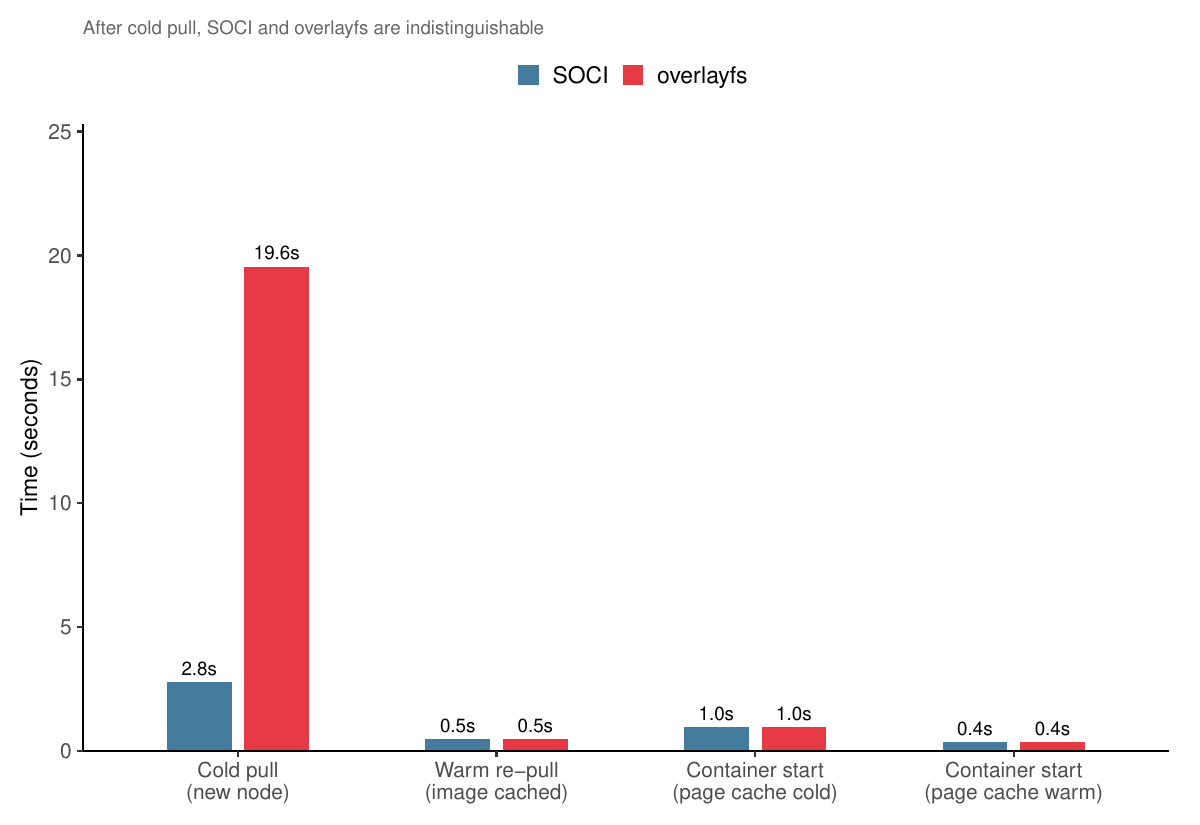}
\caption{Warm start performance is identical for SOCI and standard pull. Cold start is the only phase where they differ.}
\label{fig:warm}
\end{figure}

\subsection{Operational Considerations}

\paragraph{Index lifecycle.}
SOCI indices must be built and pushed for each image version.
In CI/CD pipelines, this is a post-push step (analogous to vulnerability scanning or signing).
Index generation takes 1--9 seconds per image (Table~\ref{tab:e5}) and can run asynchronously.
Stale indices (built for an older image version) are harmless: the snapshotter detects digest mismatch and falls back to full pull.

\paragraph{Registry cost.}
Lazy loading issues more HTTP requests than a standard pull (one range request per file access vs one GET per layer).
For ECR, range requests are not billed differently from full-layer GETs.
For registries with per-request pricing, lazy loading may increase cost for dense-access workloads.
The background prefetcher mitigates this by batching adjacent ranges.

\begin{figure}[h]
\centering
\includegraphics[width=\columnwidth]{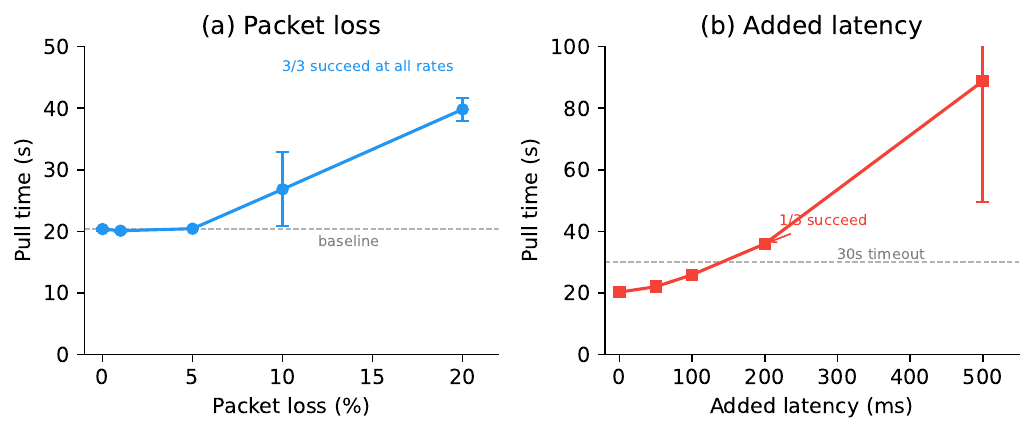}
\caption{SOCI pull behavior under network faults. Left: packet loss (all pulls succeed, time increases). Right: added latency (failures above 200ms RTT).}
\label{fig:fault}
\end{figure}
\paragraph{Failure modes.}
SOCI adds one new failure mode: the FUSE filesystem may block on a range request that times out.
We tested resilience by injecting network faults during pulls:
\begin{itemize}
\item \textbf{Packet loss (1--20\%):} All pulls succeed. TCP retransmission absorbs the loss. At 20\% loss, pull time approximately doubles.
\item \textbf{Added latency (50--500\,ms):} Pulls succeed up to 200\,ms added RTT. Beyond that, the 30\,s connection timeout triggers intermittent failures.
\item \textbf{Complete connectivity loss:} Hard failure at 30\,s. Recovery requires clearing local state and retrying from scratch; no partial-pull resume exists.
\end{itemize}
SOCI degrades gracefully under partial network degradation.
Under complete connectivity loss, the container fails after a 30\,s timeout.
This is the correct behavior: if the container cannot reach the registry to fetch file content, it cannot run, and should be rescheduled to a healthy node.

\subsection{Limitations}

\paragraph{First-access latency.}
Each file's first read requires fetching its span from the registry.
We measured first-access latency on cold spans (SOCI v0.11, ECR in-region).
The first fetch after the snapshotter starts pays connection setup cost (median 62\,ms with cached auth tokens, up to 290\,ms on a fully cold connection).
Once the HTTP connection pool is warm, subsequent fetches cost 4.6\,ms each (median, IQR 4.5--4.8\,ms), regardless of file size in the 25\,KiB--1.2\,MiB range.
100 lazy-loaded files on a warm connection add $\sim$460\,ms to container startup.
The background fetcher reduces this further by making spans local before the application reaches them.

\paragraph{FUSE overhead on cached layers.}
The decision to use FUSE is distinct from the decision to lazily load data.
SOCI only lazily loads a layer when it is not fully local at the time the container launches.
However, SOCI's mount topology uses overlayfs with per-layer FUSE mountpoints as lowerdirs, which means FUSE remains in the read path for the duration of that container's lifetime, even after the background fetcher has made all spans local.
The risk is a performance penalty for I/O-intensive operations that expect a kernel-local filesystem (metadata-heavy workloads, random small reads).

In practice, we did not detect this penalty.
4\,KiB random reads on warm-cached SOCI layers show identical latency (p50 1.85\,$\mu$s) and CPU usage to baseline overlayfs; the kernel page cache absorbs repeated reads without re-entering the FUSE userspace path.
We describe the test setup in Appendix~\ref{sec:fuse-test}.
On warm restart (all layers already in the content store from a prior pull), containerd skips the snapshotter entirely and assembles a standard overlayfs mount with no FUSE involvement.

\paragraph{Per-span integrity.}
The SOCI index stores a digest for each byte-range span. Each range response is checked against its span digest before the content reaches the running application.
The trust chain is: image manifest (layer digest, signed) $\rightarrow$ SOCI index (verified by its own digest via OCI referrers) $\rightarrow$ span digests (per-range verification on arrival).
Standard OCI pull verifies integrity only at the full-layer granularity after the entire download completes; SOCI verifies at span granularity as bytes arrive.

\paragraph{Index storage.}
Each image version requires a SOCI index (1--3\% of compressed image size).
For a registry with 1000 images averaging 1\,GB each, total index storage is approximately 10--30\,GB.
This is negligible relative to the image storage itself.

\paragraph{Compatibility.}
SOCI requires:
\begin{itemize}
\item containerd 1.7+ with remote snapshotter support
\item Registry supporting OCI referrers API (Distribution Spec 1.1)
\item Registry supporting HTTP range requests (RFC 7233)
\end{itemize}
All major cloud registries meet these requirements.
Self-hosted registries (Harbor, Nexus) support varies; SOCI falls back to full pull when requirements are not met.

\section{Conclusion}

Most containers use a fraction of their image at startup, but standard pull downloads the entire image before the container can start.
SOCI makes startup proportional to the working set by fetching content on demand.

SOCI's key innovation is an external seekable index that provides file-granularity random access over unmodified OCI images.
The image format, registry, and signatures are all unchanged.
The index exploits structural properties of gzip and tar that have been present since the format was defined.

Our evaluation shows pull-time reductions of 7--9x on cold starts (approximately 2.8\,s vs 20--25\,s), with SOCI pull time independent of image size.
SOCI lazy loading is deployed on Amazon EKS and ECS Fargate, serving production workloads since 2023. EKS Auto Mode uses SOCI's parallel pull mode for GPU instances.

Lazy loading wins below about 80\% access density. Above that, full download is marginally faster because the range-request overhead exceeds the time saved by deferring.
For dense-access workloads, parallel full pull~\cite{soci-parallel-pull-blog} is a complementary approach.
In practice, the simplest approach is to enable SOCI and disable it for workloads where it does not help.

Future work includes adaptive prefetching based on learned access patterns, span coalescing for background fetches, vulnerability scanning using only the ztoc file listing and targeted range requests (avoiding full image download), and integration with P2P distribution systems that could accelerate the range-request fetch path.

\bibliographystyle{IEEEtran}
\bibliography{references}

\clearpage
\appendices
\section{A Brief History of Seekable Containers}
\label{sec:history}

The technique underlying SOCI (random access into a gzip stream via stored decompression dictionaries) predates container lazy loading by over a decade.
This appendix traces how it got from compression theory to container systems.

\paragraph{2005: zran.c.}
Mark Adler added \texttt{zran.c} to the zlib examples directory in version 1.2.2.4 (July 2005)~\cite{zranc}.
The program demonstrates random access to a gzip stream by building an index of DEFLATE block boundaries, each recording the compressed offset and the 32\,KiB sliding-window dictionary state.
Given this index, decompression can start at any block boundary without reading prior blocks.
This is the complete theoretical foundation for what SOCI does.

\paragraph{2019: stargz.}
Brad Fitzpatrick proposed CRFS and the stargz format at Google in March 2019~\cite{crfs}.
The CRFS README states: \emph{``gzip streams are not seekable. This means that trying to read 1KB out of a file still involves pulling hundreds of gigabytes to uncompress the stream.''}
The solution: re-compress each tar entry as its own gzip member, concatenated into a valid gzip stream.
This enables seeking to file boundaries (since each file starts a new gzip member) at the cost of requiring image conversion.

The claim that ``gzip streams are not seekable'' is accurate in that gzip provides no built-in seek primitive.
But it understates what is possible with an external index.
The zran.c technique provides random access to \emph{any} point in a gzip stream, not just file boundaries, without modifying the stream.
The distinction matters: stargz requires the image producer to convert; an external index does not.

\paragraph{2020--2022: The format-conversion consensus.}
The eStargz project~\cite{estargz} refined stargz with priority-based file ordering and per-file verification.
Nydus~\cite{nydus} introduced a purpose-built filesystem image format (RAFS) with block-level deduplication.
DADI~\cite{dadi-atc20} proposed block-device-level distribution.
The OCI community discussed standardizing seekable formats (image-spec issue \#815).
A consensus emerged: lazy loading requires a new image format.

\paragraph{2022: External indexing.}
Two projects independently showed that format conversion was unnecessary.
AWS announced SOCI at re:Invent 2022 (open-sourced August 2022, first release March 2023)~\cite{soci-snapshotter}, applying the zran.c technique to build an external index over unmodified OCI layers.
Concurrently, the Nydus project added a ``zran mode'' (merged November 2022, shipped in v2.2.0 March 2023)~\cite{nydus-zran} that generates a similar external index for standard gzip layers.
Both projects arrived at the same conclusion: you do not need a new image format, you need an index that maps files to byte ranges in the existing compressed archive.

\paragraph{Why 17 years?}
The zran.c technique was available since 2005.
Container lazy loading became a research topic in 2016 (Slacker).
The gap is a framing problem: ``gzip is not seekable'' was the accepted starting point, and it directed the solution search toward format conversion.
The reframing (``gzip is seekable given an external checkpoint index'') required connecting compression internals to a container systems problem, and that connection took time to form.

\section{FUSE Overhead Test Setup}
\label{sec:fuse-test}

To measure whether FUSE adds overhead when serving fully-cached layers, we compared two configurations on the python-flask image (1.3\,GB, 10 layers) on an m5.xlarge instance.

\textbf{Config A (baseline):} Standard containerd pull with overlayfs snapshotter. All layers are local directories.

\textbf{Config B (SOCI warm cache):} SOCI pull with all spans fully cached in the local span store. Layers are still FUSE-mounted but all reads hit local cache.

We ran 10 trials of 4\,KiB random reads (fio, psync ioengine, 10\,s runtime) on a large file in a lower layer, plus sequential reads and metadata traversal (\texttt{find / -type f}).
Results: no measurable difference between configs across all workloads (p50 latency within 0.5\% for random reads, sequential reads, and metadata operations).
The SOCI snapshotter consumed fewer than 2 CPU ticks per trial when the cache was warm.

Environment: containerd 2.1.7, SOCI v0.11.0, Amazon Linux 2023, kernel 6.1.
Raw data: \texttt{experiments/e9-fuse-results.csv} in the companion repository.

\end{document}